\documentclass[lettersize,journal]{IEEEtran}
\usepackage{amsmath,amsfonts}
\usepackage{algorithmic}
\usepackage{algorithm}
\usepackage{array}
\usepackage[caption=false,font=normalsize,labelfont=sf,textfont=sf]{subfig}
\usepackage{textcomp}
\usepackage{stfloats}
\usepackage{url}
\usepackage{verbatim}
\usepackage{graphicx}
\usepackage{cite}
\usepackage{booktabs}
\usepackage{multirow}
\usepackage{array}
\usepackage{tabularx}
\usepackage{graphicx}
\usepackage{threeparttable}

\let\classAND\AND
\let\AND\relax
\usepackage{algorithmic}

\let\AND\classAND
\AtBeginEnvironment{algorithmic}{\let\AND\algoAND}

\usepackage{algorithm}
\usepackage{amsmath}
\usepackage{amssymb}
\usepackage{mathtools}
\usepackage{amsthm}
\newtheorem{theorem}{Theorem}
\newtheorem{remark}[theorem]{Remark}
\newtheorem{finding}[theorem]{Finding}

\begin{document}

\title{Can GNNs Learn Link Heuristics? A Concise Review and Evaluation of Link Prediction Methods}

\author{
Shuming Liang, Yu Ding, Zhidong Li, Bin Liang, Siqi Zhang, Yang Wang, and Fang Chen
\thanks{This paragraph of the first footnote will contain the date on which you submitted your paper for review. It will also contain support information, including sponsor and financial support acknowledgment. (Corresponding author: Shuming Liang.)}
\thanks{Shuming Liang, Zhidong Li, Bin Liang, Yang Wang, and Fang Chen are with the Faculty of Engineering and Information Technology, University of Technology Sydney, Sydney, NSW 2007, Australia (e-mail: firstname.lastname@uts.edu.au)}
\thanks{Yu Ding is with the Faculty of Engineering and Information Sciences, University of Wollongong, Wollongong, NSW 2500, Australia (e-mail: dyu@uow.edu.au).}
\thanks{Siqi Zhang is with the College of Electrical Engineering, Zhejiang University, Hangzhou, China, (e-mail: siqizhang@zju.edu.cn)}
\thanks{Manuscript received xx; revised xxx.}}

\markboth{Journal of \LaTeX\ Class Files, IEEE}%
{Liang \MakeLowercase{\textit{et al.}}: A Sample Article Using IEEEtran.cls for IEEE Journals}



\maketitle

\begin{abstract}
This paper explores the ability of Graph Neural Networks (GNNs) in learning various forms of information for link prediction, alongside a brief review of existing link prediction methods. Our analysis reveals that GNNs cannot effectively learn structural information related to the number of common neighbors between two nodes, primarily due to the nature of set-based pooling of the neighborhood aggregation scheme. Also, our extensive experiments indicate that trainable node embeddings can improve the performance of GNN-based link prediction models. Importantly, we observe that the denser the graph, the greater such the improvement. We attribute this to the characteristics of node embeddings, where the link state of each link sample could be encoded into the embeddings of nodes that are involved in the neighborhood aggregation of the two nodes in that link sample. In denser graphs, every node could have more opportunities to attend the neighborhood aggregation of other nodes and encode states of more link samples to its embedding, thus learning better node embeddings for link prediction. Lastly, we demonstrate that the insights gained from our research carry important implications in identifying the limitations of existing link prediction methods, which could guide the future development of more robust algorithms.
\end{abstract}


\section{Introduction}
\label{Introduction}

Graph Neural Networks (GNNs) have demonstrated powerful expressiveness in graph representation learning \cite{zhou2022network,khemani2024review}. However, what structural information can be learned via GNN remains an open question \cite{xu2018powerful,chen2020can,geerts2021expressiveness,bouritsas2022improving,liu2023egnn,zhiyao2024opengsl}. 
Particularly, scant attention has been directed towards this question in terms of structural information specific to two nodes. A prominent task relying on such the information is link prediction. Although a number of GNN-based link prediction models have been introduced \cite{zhang2018link,singh2021edge,zhao2022learning,guo2022linkless,wang2023multi,liu2023link,li2024evaluating,huang2024theory}, many of them lack thorough investigations into whether their models effectively learn pair-specific structural information. 

For example, SEAL \cite{zhang2018link} and its successors \cite{li2020distance,teru2020inductive,yin2022algorithm,tan2023bring} are a family of link prediction methods that attempt to use GNNs to learn pair-specific structural information represented by traditional link heuristics such as Common Neighbors, Katz index~\cite{katz1953new}, etc. SEAL \cite{zhang2018link} has proven that most link heuristics between two nodes in a graph can be computed approximately within an enclosing subgraph specifically constructed for that two nodes. 
In essence, SEAL-type methods perform link prediction by using GNNs to classify such enclosing subgraphs, with an expectation that GNNs could learn the structural information equivalent to link heuristics. However, a critical evaluation of this expectation is lacking in the literature. We demonstrate that this expectation does not hold completely. 

In this paper, we mainly study the link prediction capability of GNNs, with a focus on three aspects. 1), we explore whether GNNs can effectively learn the pair-specific structural information related to the number of common neighbors for link prediction. 
2), we present our experimental observation: incorporating trainable node embeddings can improve the performance of GNN-based link prediction models, and the denser the graph, the greater the improvement. This observation, not extensively revealed in the prior literature, has significant practical implications for selecting appropriate methods based on graph density in real-world link prediction problems. 3), we leverage insights derived from our research to provide a limitation analysis of existing link prediction methods, thereby contributing valuable perspectives for their potential improvements.

First, the majority of GNNs follow a neighborhood aggregation scheme, where each node's representation is recursively updated by aggregating the representations of that node and its neighbors \cite{corso2020principal,li2024evaluating,khemani2024review}. The learned representations are node-wise. It has been recognized that every node's representation can hardly capture information related to the number of its neighbors. This is due to the nature of the set-based pooling of the aggregation scheme, which inherently ignores the size of the neighborhood set of each node~\cite{xu2018powerful,zhang2021labeling}. 

A general strategy for applying node-wise representations learned by GNNs to downstream multiple-node tasks (e.g., link prediction, graph classification, etc.) is to combine the representations of the nodes involved in these tasks. For link prediction, we find that the combination of two nodes' representations essentially lacks the ability to capture information related to the number of common neighbors. This is mainly because node-wise representations learned by GNNs lack information about the number of neighbors of each node, and most operations of combining two nodes' representations (e.g., concatenation, Hadamard production, etc.) also do not contain any behaviors of counting how many common neighbors between two nodes. 
To empirically verify the above, we examine the link prediction performance of an approach that incorporates traditional link heuristics (e.g., Common Neighbors) into the GNN. The approach yields results either superior or comparable to those obtained by using only GNNs, experimentally supporting our analysis.


\begin{table*}[b!]
\centering

\caption{Summary of Link Prediction Information and GNN Capabilities}
\label{tab:link_prediction}
\resizebox{\textwidth}{!}{
\begin{threeparttable}

\begin{tabular}{>{\centering\arraybackslash}m{0.08\textwidth} 
                    |>{\centering\arraybackslash}m{0.25\textwidth} 
                    |>{\centering\arraybackslash}m{0.25\textwidth}
                    |>{\centering\arraybackslash}m{0.35\textwidth}}

\toprule  
\multicolumn{2}{c|}{\textbf{Forms of Information for Link Prediction}} &\textbf{Effectiveness in Link Prediction} & \textbf{Can GNNs Learn It?
} 
\\ \hline
\multirow{2}{*}{\shortstack{\\[0.5mm]Pair-specific \\structural \\information}} 
& \textbf{NCN-dependent Link Heuristics\tnote{1}}, such as Common Neighbors, Adamic-Adar, etc. (\textit{Section} \ref{sec:heuristics review})
& Highly effective in sparse graphs. Might lose discriminative power in dense graphs. (\textit{Section} \ref{sec:ncn}, \ref{sec:previousmethods})
& Limited: Struggles to capture NCN specifics, inherently washing out the explicit count of neighbors due to set-pooling aggregation. (\textit{Section} \ref{sec:GNNs}, \ref{sec:ncn})
\\ \cline{2-4} 
& \textbf{non-NCN-dependent Link Heuristics}, such as Shortest Path Distance, SimRank, etc. (\textit{Section} \ref{sec:heuristics review})
& Varied effectiveness

& Partially: Similar learning styles to certain heuristics like SimRank. (\textit{Section} \ref{sec:non-NCN-dependent})
\\ \cline{1-4} 
\multicolumn{2}{c|}{\textbf{Trainable Node Embeddings}}
& Enhances performance, especially in dense graphs. Dominant in very dense graphs (\textit{Section} \ref{sec:ncn} \ref{sec:nodeembedding}, \ref{sec:previousmethods})
 & Yes: Powerful in encoding pairwise relationships into node embeddings (\textit{Section} \ref{sec:ncn} \ref{sec:nodeembedding}, \ref{sec:previousmethods})
\\ \hline
\multicolumn{2}{c|}{\textbf{node and edge attributes}}
& Useful but sensitive to data quality (noise and missing values).
& Yes: Effectively fuse attributes through neighborhood aggregation. (\textit{Section} \ref{sec:ncn}, \ref{sec:previousmethods})
 \\ \bottomrule

\end{tabular}
\begin{tablenotes}
        \item[1] NCN refers to the Number of Common Neighbors between a pair of nodes.
\end{tablenotes}
\end{threeparttable}
}
\end{table*}


Moreover, in our experiments, we find that trainable node embeddings (different from pre-trained node embeddings, we refer to trainable node embeddings as those embeddings that can be optimized during the model training) can enhance the performance of GNN-based link prediction models, and the denser the graph, the stronger the enhancement. In particular, by only utilizing node embeddings in GCN \cite{kipf2017semi} or GAT~\cite{gat2018graph}, we are able to surpass many link prediction specific methods on two dense graphs, i.e., ogbl-ddi and ogbl-ppa \cite{hu2020open}. 
Our explanation is as follows. Compared to the model weights of a GNN that are shared across all nodes \cite{kipf2017semi,corso2020principal,khemani2024review}, each trainable node embedding is unique to its respective node. This characteristic of node embeddings can benefit the model. When the training is supervised by positive and negative link samples (i.e., two nodes are not linked), the link state of two nodes in every link sample could be encoded into the node embeddings of that two nodes and their neighboring nodes by the neighborhood aggregation algorithm of the GNN. This would enable each node embedding to remember the relationships of that node to other nodes, allowing the model to know better which two nodes are more likely to be linked or not. Moreover, in the neighborhood aggregation of the GNN, the denser graphs would allow each node to see more other nodes, leading to better learning of node embeddings for link prediction.

The insights gained in this study can help identify and interpret the limitations of existing link prediction methods, potentially directing the search for more robust algorithms. To demonstrate this, we present two case studies: first, we show that SEAL-type methods \cite{zhang2018link,teru2020inductive,yin2022algorithm,tan2023bring} could not effectively learn information about the number of common neighbors. Second, we show that NBFNet \cite{zhu2021neural} lacks the algorithmic capability to train powerful node embeddings for link prediction. Additionally, we compare the empirical performance of various link prediction methods on OGB datasets. The results can be explained with our insights.

In this paper, we begin by reviewing traditional link heuristics, which represent various types of pair-specific structural information, and examine the capacity of GNNs to learn this information. The effectiveness of node embeddings in enhancing link prediction is also explored. Section \ref{sec:heuristics review} through \ref{sec:non-NCN-dependent} delve into these aspects in detail, with key remarks and findings summarized in Table \ref{tab:link_prediction}. Section \ref{sec:previousmethods} presents an analysis of the limitations in existing methods, followed by Section \ref{sec:Implication} which discusses the implications of our findings for practical applications. Lastly, Sections \ref{sec:Limitations} and \ref{sec:Conclusion} offer a discussion on the limitations of our study and conclude the paper, respectively.

\section{Notations and Problem Definition}
Without loss of generality, we demonstrate our work on homogeneous graphs.
Let $\mathcal{G} = (\mathbb{V} , \mathbb{E}, \mathbf{X} )$ denote a graph $\mathcal{G}$ with $N$ nodes, where $\mathbb{V}$ is the set of nodes, $|\mathbb{V}|=N$, $\mathbb{E}$ is the set of edges, and $\mathbf{X}\in \mathbb{R}^{N \times f}$ is the feature matrix of nodes. The $i$-th row of $\mathbf{X}$ (i.e., $\mathbf{x}_i \in \mathbb{R}^{f}$) is the feature vector of node $i$. The adjacency matrix is $\mathbf{A}\in \mathbb{R}^{n\times n}$ in which the $i,j$-th entry (i.e., $a_{i,j}$) is $1$ if an edge exists from node $i$ to $j$ and $0$ otherwise. The degree of node $i$ is $\mathrm{deg}(i)=\sum_{j\in\mathbb{V}}a_{i,j}$. The degree of the graph $\mathcal{G}$ is the average degree of all nodes. A set of nodes connected directly to a node $v\in\mathbb{V}$ is the first-order or 1-hop neighborhood set of $v$ and is denoted by $\Gamma_v$.

Link prediction is a node-pair-specific problem, aiming to estimate the likelihood $\hat{y}_{v,u} $ of the existence of an unknown edge $\mathcal{E}_{v,u}\notin \mathbb{E}$ between two nodes $v,u\in\mathbb{V}$. Herein we refer to $v,u$ as \textit{two target nodes} in the \textit{candidate link} $\mathcal{E}_{v,u}$.

%
\section{Statistical Link Heuristics}
\label{sec:heuristics review}

In the long history of link prediction research, especially prior to the emergence of neural networks, a variety of statistical link prediction methods have been proposed \cite{liben2003link,srilatha2016similarity,kumar2020link}. These statistical methods, known as heuristic methods or link heuristics, often rely on intuitive rules or empirical observations, and often extract structural information specific to the target node pair. Therefore, we can concrete abstract pair-specific structural information using tangible link heuristics.

In this work, based on whether a link heuristic captures information related to the \textit{N}umber of \textit{C}ommon \textit{N}eighbors (\textit{NCN}) between two target nodes or not, we categorize pair-specific link heuristics (structural information) into two types: \textit{NCN-dependent} and \textit{non-NCN-dependent}. With this categorization, we offer a concise review of link heuristics in the literature, which reveals that the majority of these heuristics are NCN-dependent.

\subsection{NCN-dependent Link Heuristics}  

Table~\ref{tab:heuristic} lists several commonly-used NCN-dependent Link heuristics. As shown, Common Neighbors (CN) is defined as the size of the intersection of the first-order neighborhood sets of two nodes. Jaccard (JA) coefficient~\cite{jaccard1901etude} normalizes the CN by the size of the union of the two nodes' neighborhood sets. AdamicAdar (AA)~\cite{adamic2003friends} and Resource Allocation (RA)~\cite{zhou2009predicting} suppress the contribution of nodes by penalizing each node with its degree. Sorensen index \cite{sorensen1948method}, Salton index  \cite{salton1983introduction}, Hub Promoted index \cite{ravasz2002hierarchical}, and Hub Depressed index \cite{ravasz2002hierarchical} incorporate the degree of two target nodes with CN. We can see from Table~\ref{tab:heuristic} that these heuristics are highly dependent on the number of common neighbors between two nodes $v,u$ (i.e., $|\Gamma_v\cap\Gamma_u|$).

\begin{table}[!ht]
\caption{NCN-dependent link heuristics between nodes $v,u$.}
\label{tab:heuristic}
\begin{center}
\begin{tabular}{ll}
\toprule  
\textbf{Heuristic} & \textbf{Definition}  \\\hline
$\mathrm{CN}_{v,u}$ &$|\Gamma_v\cap\Gamma_u|$ \\  \hline
$\mathrm{JA}_{v,u}$ \cite{jaccard1901etude} &$\frac{|\Gamma_v\cap\Gamma_u|}{|\Gamma_v\cup\Gamma_u|}$ \\ \hline
$\mathrm{AA}_{v,u}$~\cite{adamic2003friends} &$\sum_{z\in\Gamma_v\cap\Gamma_u}\frac{1}{\mathrm{log}|\Gamma_z|}$ \\ \hline
$\mathrm{RA}_{v,u}$~\cite{zhou2009predicting} &$\sum_{z\in\Gamma_v\cap\Gamma_u}\frac{1}{|\Gamma_z|}$\\\hline
Sorensen index\cite{sorensen1948method} &$\frac{2|\Gamma_v\cap\Gamma_u|}{|\Gamma_v|+|\Gamma_u|}$\\\hline
Salton index \cite{salton1983introduction} & $\frac{|\Gamma_v\cap\Gamma_u|}{\sqrt{|\Gamma_v|\cdot|\Gamma_u|}}$ \\\hline
Hub Promoted index \cite{ravasz2002hierarchical} & $\frac{|\Gamma_v\cap\Gamma_u|}{min{(|\Gamma_v|,|\Gamma_u|})}$\\\hline
Hub Depressed index \cite{ravasz2002hierarchical} &$\frac{|\Gamma_v\cap\Gamma_u|}{max{(|\Gamma_v|,|\Gamma_u|})}$ \\

\bottomrule
\end{tabular}
\end{center}
\end{table}

In addition to the above, many other link heuristics are NCN-dependent. Cannistraci et al. \cite{cannistraci2013link} suggest that the likelihood of two nodes forming a link increases if their common neighbors are members of a strongly inner-linked cohort, termed local-community-links. They introduce a modified version of CN, JA, AA, and RA, denoted as CAR-based. Furthermore, some another link heuristics consider the clustering coefficient of the nodes in counting common neighbors, such as node clustering coefficient ($\sum_{z\in\Gamma_v\cap\Gamma_u|}C(z)$) \cite{wu2016link} and node-link clustering coefficient ($\sum_{z\in\Gamma_v\cap\Gamma_u|} \frac{|\Gamma_v\cap\Gamma_z|}{|\Gamma_z|-1}\times C(z) + \frac{|\Gamma_u\cap\Gamma_z|}{|\Gamma_z|-1}\times C(z)$) \cite{wu2016predicting}, where $C(z)=\frac{2|\mathcal{E}_{j,k}:j,k\in\Gamma_z, \mathcal{E}_{j,k}\in \mathbb{E}|}{|\Gamma_z|\times (|\Gamma_z|-1)}$ is the local clustering coefficient of the node $z$.

An important family of link heuristics is those counting all paths between two target nodes, such as Katz index~\cite{katz1953new}, Leicht Holme Newman index \cite{leicht2006vertex}, and Rooted PageRank \cite{zhang2018link}. These heuristics are indeed NCN-dependent, where first-order and high-order common neighbors are considered. For example, Katz index ($\mathrm{Katz}_{v,u}=\sum_{l=1}^{\infty}\beta^l|\{\mathrm{path}_{v,u}^{(l)}\}|$) \cite{katz1953new} weighted sums the number of all paths between two target nodes $v,u$, where $|\{\mathrm{path}_{v,u}^{(l)}\}|$ is the number of all paths between node $v$ and $u$ with the length of $l$, and $\beta$ is a damping factor. For example, if $l=4$, $|\{\mathrm{path}_{v,u}^{(4)}\}|$ can be computed by $|\{\mathrm{path}_{v,u}^{(4)}\}| = \sum_{a\in\Gamma_v,b\in\Gamma_u} \mathrm{CN}_{a,b}$, where $\mathrm{CN}_{a,b}$ is the number of common neighbors between nodes $a,b$.

\subsection{non-NCN-dependent Link Heuristics} 
\label{sec:non-ncn review}
According to our categorization strategy, there exist a limited number of link heuristics that are non-NCN-dependent, including Shortest Path Distance (SPD), Preferential Attachment ($\mathrm{PA}_{v,u}=|\Gamma_v|\times |\Gamma_u|$) \cite{barabasi2002evolution}, and SimRank \cite{jeh2002simrank}. Notably, SPD and PA do not extract information about the number of common neighbors. SimRank, detailed in Algorithm~\ref{alg:simrank}, recursively refines similarity scores between every two nodes by considering the neighboring nodes of the two nodes, where the number of common neighbors is ignored essentially. Specifically, the similarity score $s_{i,j}^{(m)}$ between node $i, j$ in the $m$-th iteration is obtained by averaging the similarity scores between all neighbors of $i$ and $j$ from the $(m-1)$-th iteration, where the information about how many common neighbors between $i, j$ can hardly be encoded into $s_{i,j}^{(m)}$.
%
\begin{algorithm}[H]
   \caption{SimRank \cite{jeh2002simrank}}  
   \label{alg:simrank}         
    \begin{algorithmic}[1]
       \STATE {\bfseries Input:} Graph $\mathcal{G} = (\mathbb{V} , \mathbb{E} )$ ($|\mathbb{V}|=N$), decay factor $C$ ($0<C<1$), iterations $K$
       \STATE {\bfseries Output:} Similarity $\mathbf{S} = (s_{i,j})\in\mathbb{R}^{N\times N}$
       \STATE {\bfseries Initialize:} $s_{i,j}^{(0)} = 1$ if $i=j$, otherwise $0$
       \FOR{$m=1$ {\bfseries to} $K$}
       \STATE $s_{ij}^{(m)} = \frac{C}{|\Gamma_i||\Gamma_j|}\sum_{b=1}^{|\Gamma_j|}\sum_{a=1}^{|\Gamma_i|}s^{(m-1)}_{\Gamma_i(a)\Gamma_j(b)}$, where $\Gamma_i(a)$ is the $a$-th node in $\Gamma_i$                  
      \ENDFOR
    \end{algorithmic}
\end{algorithm}

\section{Aggregation-based GNNs}
\label{sec:GNNs}
Most GNNs follow a neighborhood information aggregation algorithm, where the representation of each node in a graph is iteratively updated by aggregating the representations of its neighbors and its own \cite{corso2020principal,khemani2024review}. Formally, the representation of a node $i$ updated by the $l$-th layer of a GNN is
\begin{equation}\label{eq:agg_gnn}
\begin{aligned}
\hat{\mathbf{h}}^{(l)}_{i} &= \mathrm{AGG}^{(l)} \left(\left\{ \mathbf{h}^{(l-1)}_{j}~|~\forall j\in \Gamma_i\cup\{i\}\right\} \right),\\
\mathbf{h}^{(l)}_{i} &= \hat{\mathbf{h}}^{(l)}_{i}\mathbf{W}^{(l)},
\end{aligned}
\end{equation}
where $\mathbf{h}^{(0)}_{i}$ is initialized with the feature vector of node $i$, $\mathrm{AGG}^{(l)}(\cdot)$ is instantiated as a set-based pooling operation such as MAX, MEAN~\cite{sage2017inductive}, or attention-based SUM~\cite{gat2018graph,li2024evaluating}, $\mathbf{W}^{(l)}$ is a weight matrix for the $l$-th GNN layer, which is shared across all nodes and used for representation transformation (i.e., if $\hat{\mathbf{h}}^{(l)}_{v}\in\mathbb{R}^{f}, \mathbf{W}^{(l)}\in\mathbb{R}^{f\times f'}$, then $\mathbf{h}^{(l)}_{v}\in\mathbb{R}^{f'}$).
For simplicity, we omit the residual connections, activation functions, etc. In this paper, we use the term GNNs to refer to such aggregation-based GNNs unless otherwise stated.

\begin{figure}[!ht]
\begin{center}
\centerline{\includegraphics[width=0.75\columnwidth]{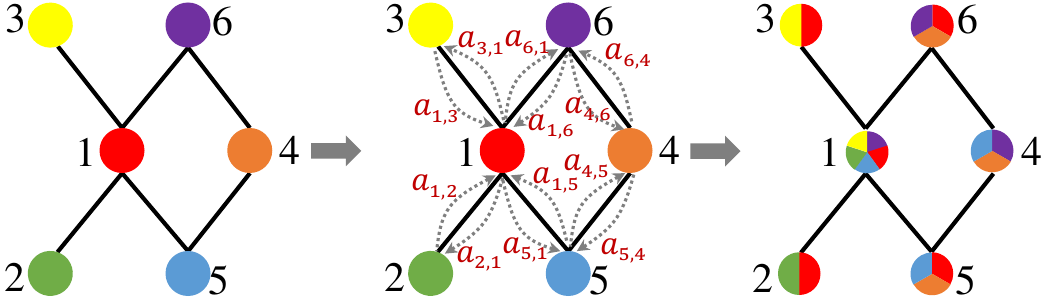}}
\caption{An illustration of neighborhood information propagation and aggregation in GNNs, where $a_{i,j}$ can be an edge weight or attention weight from node $j$ to $i$.}
\label{fig:GNN}
\end{center}
\end{figure}

Fig. \ref{fig:GNN} illustrates the neighborhood information propagation and aggregation process in GNNs. As shown, $\hat{\mathbf{h}}^{(l)}_{i}$ in Eq. \ref{eq:agg_gnn} can be computed by
\begin{equation}\label{eq:gcn_gat}
\hat{\mathbf{h}}^{(l)}_{i} = \sum_{j\in\Gamma_i\cup\{i\}}\frac{a_{i,j}^{(l)}}{\sum_{j\in\Gamma_i\cup\{i\}}a_{i,j}^{(l)}}\mathbf{h}^{(l-1)}_{j},
\end{equation}
where $a_{i,j}^{(l)}$ is the weight for the message (i.e., $\mathbf{h}^{(l-1)}_{j}$) from node $j$ to $i$. For MEAN-pooling in GNNs like GCN~\cite{sage2017inductive}, it can be $a_{i,j}^{(l)}=1, \forall \mathcal{E}_{i,j}\in \mathbb{E}$. For attention-based GNNs like GAT~\cite{gat2018graph}, $a_{i,j}^{(l)}$ is an attention coefficient that is computed dynamically based on $\mathbf{h}^{(l-1)}_{i}$ and $\mathbf{h}^{(l-1)}_{j}$.

\begin{remark}
\label{prop:nodewise}
The node representations learned by aggregation-based GNNs are node-wise.
\end{remark}

\noindent\textbf{Analysis.} As shown in Eq.~\ref{eq:agg_gnn}, the input, intermediate, and output representations of GNNs are node-wise. 

\begin{remark}\label{prop:prop:gnn_num}
Each node's representation learned by neighborhood aggregation-based GNNs lacks information about the number of neighboring nodes of that node.
\end{remark}
\noindent\textbf{Analysis.} 
As shown in Eq.~\ref{eq:agg_gnn}, GNNs update the representation $\mathbf{h}^{(l)}_{i}$  by aggregating the representations of node $i$ and its neighbors. In this process, the number of neighbors of node $i$ can hardly be encoded into $\mathbf{h}^{(l)}_{i}$. This is due to the nature of the neighborhood aggregation scheme in GNNs, i.e., the $\mathrm{AGG}^{(l)}(\cdot)$ in Eq.~\ref{eq:agg_gnn} is set-based pooling, which is originally designed to handle irregular sizes of neighborhood sets of different nodes in a graph. For example, if the aggregation is MEAN pooling, then the set of node-wise representations (i.e., $\{ \mathbf{h}^{(l-1)}_{j}~|~\forall j\in \Gamma_i\cup\{i\}\}$ in Eq.~\ref{eq:agg_gnn}) will be averaged and the result could hardly contain information about the size of that set. 
Note that attention-based pooling also cannot address this inherent issue of the neighborhood aggregation scheme. As shown in Eq.~\ref{eq:gcn_gat}, attention-based GNNs~\cite{gat2018graph} essentially replace the original edge weight with attention weight. Despite this modification, the set of representations is weighted averaged, and consequently, the resulting representation still lacks information related to the size of the neighborhood set. 

Remark~\ref{prop:prop:gnn_num} shows that the neighborhood aggregation algorithm of GNNs cannot effectively learn information about the number of neighbors of each node. Essentially, we can address this issue by, for example, adding the node degree as a feature to each node. We note that several previous works have pointed out this \cite{xu2018powerful,zhang2021labeling}. We present it using Remark~\ref{prop:prop:gnn_num} for better presenting our following analysis.

\section{Can GNNs Completely Learn NCN-dependent Structural Information?}
\label{sec:ncn}
%
\subsection{Analytical Study}
What can we do when applying the node-wise representations learned by GNNs to downstream graph tasks that involve multiple nodes, such as link prediction or graph classification? A general way is to combine the representations of the involved nodes into one representation and pass it into the subsequent model components \cite{xu2018powerful,wang2021pairwise,li2024evaluating,khemani2024review}. For such a combination, we have the following insight:
\begin{remark}\label{prop:GNN_cn}
The combination of two or more nodes' representations learned by GNNs cannot effectively capture NCN-dependent structural information.
\end{remark} 

\noindent\textbf{Analysis.} According to Remark~\ref{prop:prop:gnn_num}, due to the nature of the neighborhood aggregation algorithm, node-wise representations learned by GNNs cannot effectively capture information related to the number of neighbors of each node (i.e., the size of its neighborhood set), much less to the number of common neighbors between two nodes. The operation of combining representations of two or more nodes also cannot effectively extract NCN-dependent structural information. For example, we can combine the representations of two nodes by concatenation, Hadamard production \cite{wang2021pairwise,khemani2024review}, and combine more nodes' representations by MEAN pooling \cite{xu2018powerful}, Sort pooling \cite{zhang2018end}, etc. These combination operations on node-wise representations learned by GNNs are unlikely to contain the behavior of counting the common neighbors between two nodes, thereby falling short in extracting NCN-dependent structural information.

GNNs might learn little structural information related to the number of common neighbors. However, the neighborhood aggregation algorithm of GNNs learns node-wise representations by passing the messages of neighboring nodes of each node to that node and set-based aggregates them \cite{sage2017inductive,gat2018graph,zhang2021labeling,li2024evaluating}. Such set-based aggregation operation inherently washes out the information related to the number of nodes in the set, including the number of common nodes between two nodes. For instance, as shown in Fig.~\ref{fig:GNN}, the nodes $1$ and $4$ have common neighboring nodes $5,6$. In GNN learning based on Eq. \ref{eq:agg_gnn}, the node $1$ will receive the messages from $2,3,5,6$, where the number of common neighbors (i.e., $\mathrm{CN}_{1,4}=2$) can hardly be captured in the aggregation of five representations (i.e., representations of nodes $1,2,3,4,5$). Note that in this example, attention-based aggregation also cannot effectively learn $\mathrm{CN}_{1,4}=2$ from the aggregation of five representations. The reason is the same as the analysis of Remark~\ref{prop:prop:gnn_num}. In fact, it is difficult to interpret what information the GNN has learned in a rigorous mathematical format (an acknowledged limitation of our study (see Section~\ref{sec:Limitations})). Nevertheless, we could say that GNNs cannot completely capture NCN-dependent structural information.

\begin{algorithm}[tb]
   \caption{Link prediction by integrating statistical heuristics into the GNN}
    \begin{algorithmic}[1]
   \label{alg:GNNLP}
   \STATE {\bfseries Input:} Graph $\mathcal{G} = (\mathbb{V} , \mathbb{E}, \mathbf{X} )$ ($|\mathbb{V}|=N$), $\mathbf{X}\in\mathbb{R}^{N\times f}$, trainable node embeddings $\mathbf{E}\in\mathbb{R}^{N\times d}$, trainable heuristic embeddings, ground truth $y_{v,u}$ for link sample $(v,u)$, GNN layers $L$, epochs $K$
   \STATE {\bfseries Output:} Link likelihood $\hat{y}_{v,u}\in\mathbb{R}$ for node pair $v,u$
   \STATE {\bfseries Initialize:} node embeddings $\mathbf{E}$, trainable heuristic embeddings, model weights, etc.
   \FOR{$i=0$ {\bfseries to} $K$}
       \FOR{$l=1${\bfseries to} $L$}
           \STATE $\hat{\mathbf{h}}^{(l)}_{i} = \mathrm{AGG}^{(l)} \left(\left\{ \mathbf{h}^{(l-1)}_{j}~|~\forall j\in \Gamma_i\cup\{i\}\right\} \right)$
           \STATE $\mathbf{h}^{(l)}_{i} = \hat{\mathbf{h}}^{(l)}_{i}\mathbf{W}^{(l)}$
       \ENDFOR
       \STATE $\mathbf{h}_{vu} =\mathrm{COMBINE}\left(\mathbf{h}^{(L)}_{v},\mathbf{h}^{(L)}_{u}\right)$
       \STATE $\mathbf{e}_{vu} = \mathrm{CONCAT} \left(\mathbf{e}_{vu}^{(\mathrm{CN})},\mathbf{e}_{vu}^{\mathrm{(JA)}},\cdots,\mathbf{e}_{vu}^{\mathrm{(RA)}}\right)$
       \STATE $\hat{y}_{vu} = \mathrm{PREDICTOR}\left(\mathrm{CONCAT}\left(\mathbf{h}_{vu},\mathbf{e}_{vu}\right)\right)$
       \STATE Calculate $\mathrm{loss}(y_{v,u}, \hat{y}_{v,u})$
       \STATE Update $\mathbf{E}$, trainable heuristic embeddings, model weights, etc.
   \ENDFOR
   \STATE Herein $\mathbf{h}^{(0)}_{i}$ is initialized based on the feature and node embedding of node $i$ (i.e.,  $\mathbf{x}_{i}$ and $\mathbf{e}_{i}$). $\mathbf{h}_{vu}$ is the link representation for $(v,u)$. $\mathbf{e}_{vu}^{(\mathrm{CN})},\mathbf{e}_{vu}^{\mathrm{(JA)}},\mathbf{e}_{vu}^{\mathrm{(RA)}}$ are trainable heuristic embeddings by encoding CN, JA, RA between nodes $v,u$, respectively. $\mathrm{COMBINE}(\cdot,\cdot)$ can be Hadamard production, concatenation, etc. $\mathrm{CONCAT}$ is the operation of concatenation. $\mathrm{PREDICTOR}(\cdot)$ is a predictor like MLP.
   
\end{algorithmic}
\end{algorithm}

\subsection{Empirical Study} 
\subsubsection{Experimental Design}

If Remark~\ref{prop:GNN_cn} holds, we expect that properly integrating NCN-dependent heuristics with a GNN could improve the link prediction performance. To this end, we design our experiments as detailed in Algorithm~\ref{alg:GNNLP}. As shown, given two nodes $v,u$, the link prediction is performed by combining the two nodes' representations from the last GNN layer into a pair-specific link representation $\mathbf{h}_{vu}$, then concatenating it with heuristic encodings, and lastly passing the concatenation into a predictor like MLP. During the training stage, the node pair $(v,u)$ can be a positive or negative link sample, where a negative sample can be two distant nodes that are not connected to each other.

In Algorithm~\ref{alg:GNNLP}, $\mathbf{h}^{(0)}_{i}$ can be initialized using feature vector $\mathbf{x}_i\in \mathbf{X}$, node embedding $\mathbf{e}_i\in \mathbf{E}$ or the concatenation of both $\mathbf{x}_i$ and $\mathbf{e}_i$. Some may refer to node embedding as an intermediate representation of a node in GNNs. In this work, we clearly distinguish node embeddings from node representations. We consider node embedding as a type of node-wise input feature. The embedding of a node can be viewed as encoding a unique node id into a trainable embedding vector, which is like encoding a unique word id into the word embedding in natural language processing \cite{mikolov2013distributed}. Note that we can encode any feature into a trainable embedding vector (e.g., encoding node degree to an embedding). The main difference between node embeddings and the embeddings of other node features is that each node embedding vector is unique to that node. For example, an embedding of a node degree is not unique to that node (different nodes could have the same node degree). 

We also encode link heuristics into trainable embedding vectors. To verify Remark~\ref{prop:GNN_cn}, we only need to encode NCN-dependent link heuristics. The methodology of encoding heuristics is as follows. For heuristics that are discrete integer values (e.g., CN), we assign a trainable embedding vector to each integer. In the case of heuristics that are continuous floating-point values (e.g., AA), we partition the value range into small bins and subsequently allocate each bin a unique embedding vector. Encoding heuristics into embeddings is mainly because if we directly use heuristics as features, we find that the model optimization is challenging, where the model is more likely to get stuck in a local optimum. This issue could arise due to the high correlation between the heuristic features and the link samples. Encoding heuristics into trainable embeddings can address this issue successfully.

\subsubsection{Datasets}

\begin{table}[th!]
\caption{Statistics of OGB link prediction datasets used in our experiments.}\label{tab:datasets}
\begin{center}
\begin{sc}
\begin{tabular}{lrrr}
\toprule 
\textbf{Dataset} & \textbf{\#Nodes} & \textbf{\#Edges}& \textbf{\#Degree} \\\midrule
ogbl-ddi &$4,267$&$1,334,889$&$500$ \\  \midrule
ogbl-collab &$235,868$&$1,285,465$& $8$\\  \midrule
ogbl-ppa &$576,289$&$30,326,273$& $73$\\ \midrule
ogbl-citation2 &$2,927,963$ &$30,561,187$&$21$\\\bottomrule
\end{tabular}
\end{sc}
\end{center}
\end{table}

All of our experiments are conducted on four OGB link prediction datasets: \textit{ogbl-collab, ogbl-citation2, ogbl-ppa}, and \textit{ogbl-ddi}~\cite{hu2020open}. The statistics of datasets are summarized in Table~\ref{tab:datasets}.
All these datasets are constructed based on real-world data, covering diverse realistic applications and spanning different scales (4K - 3M nodes). 
OGB provides an official evaluation protocol.
We completely follow it in the data splits and evaluation metrics (i.e., Hits@$50$, MRR, Hits@$100$, and Hits@$20$ on ogbl-collab, ogbl-citation2, ogbl-ppa and ogbl-ddi, respectively). 
We report the result on the test set, with mean and standard deviation computed across 10 trials.

\begin{figure*}[ht!]
\begin{center}
\centerline{\includegraphics[width=0.8\textwidth]{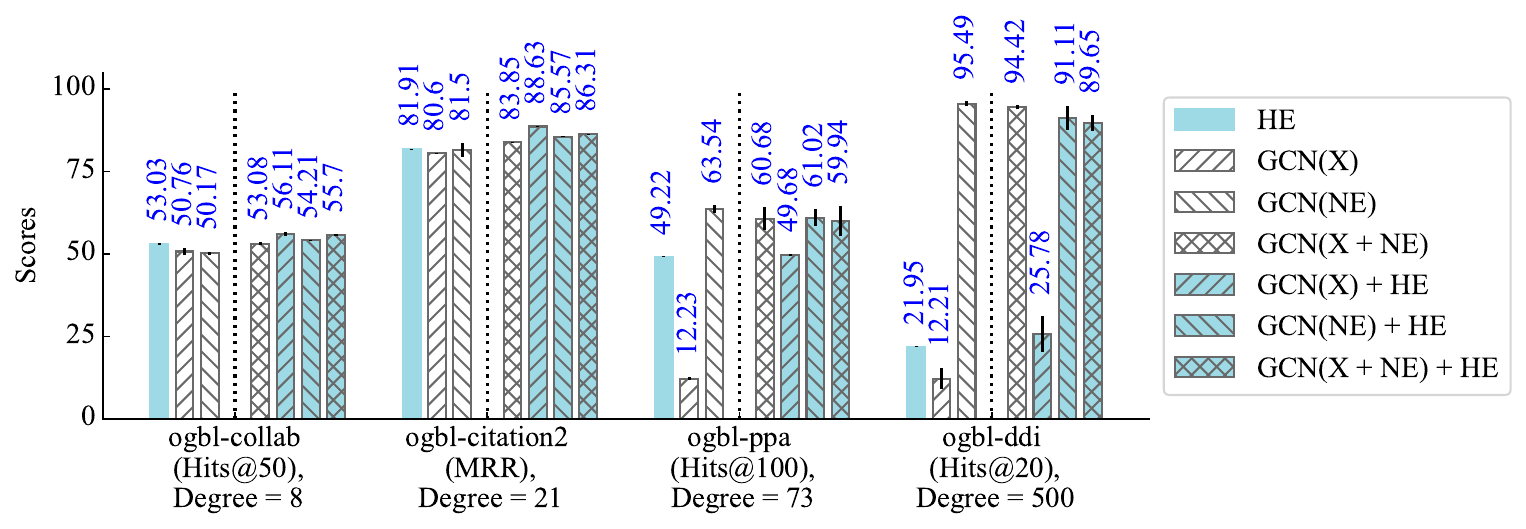}}
\caption{The results of Algorithm~\ref{alg:GNNLP} on four OGB link prediction datasets, using heuristic encoding (HE) only, node features (X) only, node embeddings (NE) only, or their combinations. The data splits and evaluation metrics follow OGB official evaluation protocol \cite{hu2020open}.
}
\label{fig:hexne}
\end{center}
\end{figure*}

\subsubsection{Implementation Details}\label{sec:Implementation}

We implement our Algorithm~\ref{alg:GNNLP} based on PyTorch and PyTorch Geometric~\cite{FeyLenssen2019}. All embedding vectors are initialized by following the methods in \cite{glorot2010understanding,he2015delving}. The model is trained with Adam optimizer~\cite{kingma2014adam}. The learning rate is decayed using the ExponentialLR method \cite{Li2020An}. 
We conduct experiments on ogbl-ddi and ogbl-collab on a Linux machine with 192G RAM, and NVIDIA Quadro P6000 (24G), and on ogbl-ppa and ogbl-citation2 on a machine with 512G RAM and NVIDIA A100 (40G). 
Table~\ref{tab:Implementation} lists the configurations of the Algorithm~\ref{alg:GNNLP} for the best performance. We provide our code for reproducing the results at \textit{https://github.com/astroming/GNNHE}.
\begin{table}[ht!]
    \centering
    \caption{Configurations of the Algorithm~\ref{alg:GNNLP} for the best performances.}
    \label{tab:Implementation}
    \scalebox{0.89}{
    \begin{tabular}{lcccc}
        \toprule
        & \multicolumn{1}{c}{\textbf{ogbl-ddi}} & \multicolumn{1}{c}{\textbf{ogbl-collab} } & \multicolumn{1}{c}{\textbf{ogbl-ppa}} &\multicolumn{1}{c}{\textbf{ogbl-citation2}}        
        \\
        \midrule
        GNN module &GCN&GCN&GCN&GCN
        \\
        GNN layers &2&2&2&2
        \\
        predictor &MLP&MLP&MLP&MLP
        \\
        predictor layer&4&5&3&4
        \\
        heuristics &-&SPD,CN,AA&-&SPD,AA
        \\
        heuristic embedding dim &-&32&-&32
        \\
        node embedding dim &512&-&256&-
        \\
        lr & 0.003&0.002&0.001&0.001
        \\
        dropout rate &0.3&0.3&0.3&0.25
        \\
        gradient clip norm &5&10&5&10
        \\
        batch size & 100000&70000&100000&15000
        \\
        \bottomrule\\[-8pt]
    \end{tabular}
    }
\end{table}

\subsubsection{Experimental Results}
Fig. \ref{fig:hexne} shows the experimental results, where HE is the model that only uses the heuristic encoding $\mathbf{e}_{vu}$ in Line 11 of Algorithm~\ref{alg:GNNLP}.  
GNN(X), GNN(NE), and GNN(X+NE) denote the models only using the GNN with three different inputs: node features (X) only, node embeddings (NE) only, and concatenation of both X and NE (X+NE). The model GNN(X)+HE uses both $\mathbf{h}_{vu}$ and $\mathbf{e}_{vu}$. 

We can see in Fig.~\ref{fig:hexne} that HE outperforms GNN(X) on all datasets, suggesting that NCN-dependent heuristics convey meaningful information that could not be effectively learned by the GNN. Moreover, most of the results of combining GNN and HE are better than those only using GNN. Especially, GNN(X)+HE achieves the best on ogbl-collab and ogbl-ciation2. All these results can support Remark~\ref{prop:GNN_cn}.

\section{GNNs with Node Embedding in Link Prediction}
\label{sec:nodeembedding}
As shown in Fig.~\ref{fig:hexne}, on two relatively sparse graphs, i.e., ogbl-collab (degree 8) and ogbl-citation2 (degree 21), the performance of GNN(NE) is on par with that of GNN(X), and GNN(X+NE) performs better than GNN(NE) and GNN(X). By comparison, on two denser graphs, i.e., ogbl-ppa (degree 73) and ogbl-ddi (degree 500), GNN(NE) outperforms GNN(X) by a large margin. These results indicate that incorporating node embeddings into GNNs can enhance the link prediction performance. More importantly, it reveals a strong positive correlation between the graph degree and the performance improvement by node embeddings, i.e., denser graphs exhibit greater improvement.

\begin{remark}\label{prop:gnn_ne_strong}
For GNN-based link prediction models like Algorithm~\ref{alg:GNNLP}, when the training is supervised by positive and negative links, trainable node embeddings could enhance the expressive power of these models.
\end{remark}
\noindent\textbf{Analysis.}
In Algorithm~\ref{alg:GNNLP}, the parameters optimized by the link samples could include model weights, trainable node embeddings, and other embedding weights. The key difference between node embedding weights and other learnable weights is that the former is unique to each node but the latter is shared across multiple nodes (e.g., the GNN weight matrix $\mathbf{W}^{(l)}$ in Eq.~\ref{eq:agg_gnn} is shared across all nodes). The unique nature of node embeddings can bring benefits. As shown in Algorithm~\ref{alg:GNNLP}, when the model training is supervised by a link sample $(v,u)$, for a GNN using node embeddings, the loss calculated based on $(y_{v,u}, \hat{y}_{v,u})$ would be used to optimize the node embeddings of nodes $v,u$ and their neighboring nodes (i.e., the nodes involved in calculating $\mathbf{h}^{(L)}_{v},\mathbf{h}^{(L)}_{u}$). The link state of $(v,u)$ could be encoded into the node embeddings of these nodes, which would enable those node embeddings to remember the relationships between corresponding nodes. After being trained with sufficient positive and negative link samples, the node embedding of each node could know which nodes (through their node embeddings) in the graph are more likely to be or not to be connected to that node. 

If node embeddings are not used in Algorithm~\ref{alg:GNNLP}, link samples will only supervise the optimization of the weights that are shared across multiple nodes. The states of link samples could not be effectively preserved by the model since these shared weights might learn a common pattern for different nodes rather than unique to a node. By comparison, each node embedding is unique to that node and could learn the link information specific to that node. In this respect, trainable node embeddings could enhance the expressive power of the GNN-based link prediction model.

In Remark~\ref{prop:gnn_ne_strong}, the requirement of \textit{the model training is supervised by positive and negative links} is indispensable. Without this prerequisite, the link state between two nodes could not be encoded into node embeddings. Additionally, negative link samples can allow the embeddings of two distant nodes and their neighbors to see each other during the optimization of the GNN-based model. 

\begin{finding}\label{finding:denser}
Following Remark~\ref{prop:gnn_ne_strong}, the denser the graph, the more the enhancement by node embeddings.
\end{finding}
\noindent\textbf{Analysis.} 
In GNN-based link prediction models like Algorithm~\ref{alg:GNNLP}, node embeddings in a dense graph could be better learned for link prediction than those in a sparse graph. Our explanation is as follows. In a dense graph, a node often has a lot of neighboring nodes, thereby providing numerous opportunities for that node to meet other nodes and encode link relationships of that node with these other nodes into its embedding during the GNN training. In contrast, a sparse graph typically contains only a limited number of neighbors for each node. For example, in the case where a node $v$ has only one neighboring node $w$, the optimization of the embedding of node $v$ in a GNN would mainly rely on its neighbor $w$. As a result, the learned embedding of node $v$ would lack sufficient information to identify the relationships between node $v$ and the majority of the other nodes in the sparse graph because node $v$ rarely or never sees them during the training process.

Finding~\ref{finding:denser} indicates that the graph degree significantly influences the effectiveness of trainable node embeddings in GNN-based link prediction models. Interestingly, prior studies \cite{shang2019link} have also highlighted the sensitivity of heuristic methods to the graph degree. This underscores the necessity of considering the graph degree when selecting link prediction methods, as their efficacy may vary depending on it. Investigating the influence of different graph degrees on link prediction methods represents a compelling direction for further research.

\section{GNNs in Learning non-NCN-dependent Structural Information}
\label{sec:non-NCN-dependent}
In Section \ref{sec:ncn}, we present that GNNs cannot completely learn the structural information related to the number of common neighbors between two target nodes. When it comes to the question of what non-NCN-dependent pair-specific structural information can be learned via GNNs, it poses a significant challenge. This is due to the potential presence of diverse types of non-NCN-dependent information. Unlike NCN-dependent information directly related to the number of common neighbors, non-NCN-dependent information tends to be abstract and difficult to express in rigorous mathematical terms. For example, PA relies on the product of node degrees, reflecting a commonly observed phenomenon in social and web networks where newly added nodes are more likely to connect to nodes with high degrees. Our review in Section \ref{sec:non-ncn review} highlights three widely used non-NCN-dependent link heuristics. In this work, we leave the exploration of this question as a future research endeavor. Nevertheless, by comparing Algorithm~\ref{alg:simrank} and Algorithm~\ref{alg:GNNLP}, we have the following insight.

\begin{remark}\label{remark:gnnsimrank}
The learning styles of SimRank in Algorithm~\ref{alg:simrank}, and the GNN-based link prediction model with node embeddings in Algorithm~\ref{alg:GNNLP}, exhibit certain similarities.\footnote{For Remark~\ref{remark:gnnsimrank}, we do not compare the performance of Algorithm~\ref{alg:simrank} and Algorithm~\ref{alg:GNNLP} due to the difficulty of SimRank in computation. For example, the basic memory requirement of SimRank is 415G and 2.42T on ogbl-collab and ogbl-ppa, respectively. Besides, SimRank produces 0 at Hits@20 on ogbl-ddi.}
\end{remark}

\noindent\textbf{Analysis.} 
Comparing Algorithms~\ref{alg:simrank} and \ref{alg:GNNLP}, several similarities emerge. Firstly, similarity scores in Line 3 of Algorithm~\ref{alg:simrank} and node embeddings in Line 3 of Algorithm~\ref{alg:GNNLP} both need to be initialized and can be dynamically trained. Secondly, the updating computations of both algorithms (i.e., Line 5 of Algorithm~\ref{alg:simrank} and Line 6 of Algorithm~\ref{alg:GNNLP}) involve neighboring nodes. Moreover, both the learned results (i.e., $s_{ij}$ in Algorithm~\ref{alg:simrank} and $\hat{y}_{v,u}$ in Algorithm~\ref{alg:GNNLP}) describe the existence likelihood of a link between two nodes. However, compared to Algorithm~\ref{alg:GNNLP} where the trainable parameters include node embeddings, model weights, etc., the expressive power of SimRank is limited. In SimRank, only the similarity scores between every two nodes can be optimized, with each score always taking the form of a scalar. 

Remark~\ref{remark:gnnsimrank} implies that although NCN-dependent structural information cannot be effectively learned via GNNs (Remark~\ref{prop:GNN_cn}), other types of information (e.g., the information captured by SimRank) might be learned through GNNs.

\section{Limitation Analysis of Existing Methods}\label{sec:previousmethods}
In this section, we first present a brief survey of existing link prediction methods and then identify their possible limitations.

\subsection{A Survey of Link Prediction Methods}\label{sec:review}

\subsubsection{Heuristic Methods} 
As outlined in Section \ref{sec:heuristics review}, traditional link heuristics are usually defined based on the number of common neighbors or paths between two nodes~\cite{martinez2016survey}. Their effectiveness in link prediction has been confirmed in real-world tasks \cite{martinez2016survey,kovacs2019network}. However, many link heuristics are designed for specific graph applications and their performance may vary on different graphs \cite{kovacs2019network}. Also, the expressiveness of these methods is limited compared to graph representation learning methods~\cite{zhang2018link}.

\subsubsection{Graph Neural Networks}

GNNs have proven their effectiveness in various graph applications \cite{liang2021failure,huang2020skipgnn,ying2021transformers,khemani2024review}.
A number of GNN models have been proposed \cite{kipf2017semi,xu2018representation,zou2023simple}.
GCN ~\cite{kipf2017semi} learns node representations by summing the normalized representations from the first-order neighbors.
GraphSAGE~\cite{sage2017inductive} samples and aggregates representations from local neighborhoods. 
GAT~\cite{gat2018graph} introduces an attention-based GNN architecture. JKNet \cite{xu2018representation} adds a pooling layer following the last GNN layer and each GNN layer has a residual connection to this layer. Cluster-GCN \cite{chiang2019cluster} proposes an efficient algorithm for training deep GCN on large graphs. 
LRGA~\cite{puny2020global} incorporates a Low-Rank global attention module to GNNs. 
Several works such as Mixhop \cite{abu2019mixhop} and DEGNN \cite{li2020distance} propose techniques to leverage high-hop neighbors. ID-GNN \cite{you2021identity} embeds each node by considering its identity. 
These GNNs have demonstrated promising link prediction performance. 

\subsubsection{Non-GNN-based Node Embedding Methods}
A family of node embedding methods is those built with matrix factorization \cite{koren2009matrix}. MF \cite{menon2011link} is a pioneer work employing matrix factorization in link prediction. FSSDNMF \cite{chen2022link} proposes a link prediction model based on non-negative matrix factorization. In general, such methods mainly rely on the adjacency matrix and tend to encounter scalability issues when employed on large graphs.

Another family of node embedding methods is those based on relative distance encoding. 
The similarity of nodes in the embedding space reflects the semantic similarity of nodes in the graph~\cite{perozzi2014deepwalk}. Such methods would learn more similar embeddings for two close nodes than two distant nodes. 
Following word embedding~\cite{mikolov2013distributed}, methods such as Deepwalk~\cite{perozzi2014deepwalk}, Node2vec~\cite{grover2016node2vec}, and NodePiece~\cite{galkin2021nodepiece} learn node embeddings by treating the nodes as words and treating the sequences of nodes generated based on links as sentences. UniNet \cite{yao2021uninet} improves the efficiency of such methods using the Metropolis Hastings sampling technique \cite{chib1995understanding}. 
Inspired by subword tokenization~\cite{sennrich2016neural}, NodePiece~\cite{galkin2021nodepiece} explores parameter-efficient node embeddings. 

\subsubsection{SEAL-type Methods}
SEAL-type methods have shown superior performance among existing link prediction approaches \cite{hu2020open,li2024evaluating,li2024evaluating}.
SEAL and its subsequent works \cite{zhang2018link,li2020distance,teru2020inductive,yin2022algorithm,tan2023bring} address the link prediction problem by classifying the subgraphs that are extracted specifically for candidate links. SEAL~\cite{zhang2018link} extracts a local enclosing subgraph for each candidate link and uses a GNN~\cite{zhang2018end} to classify these subgraphs for link prediction. GraiL~\cite{teru2020inductive} is developed for inductive link prediction. It is similar to SEAL but it replaces SortPooling \cite{zhang2018end} with MEAN-pooling. 
DEGNN~\cite{li2020distance} proposes a distance encoding GNN. 
Cai et al. \cite{cai2021line} transform the enclosing subgraph into a corresponding line graph and address the link prediction task with the node classification problem in its line graph. 
Pan et al.\cite{pan2021neural} follow the subgraph strategy in SEAL while designing a new pooling mechanism called WalkPool. 
SUREL~\cite{yin2022algorithm} proposes an algorithmic technique to improve the computational efficiency of subgraph generation in SEAL. 
SIEG \cite{ai2022structure} incorporates the structural information learned from the enclosing subgraphs into the GNN for link prediction, which fuses topological structures and node features to take full advantage of graph information for link prediction.  

\subsubsection{Methods Specific for Link Prediction}
Various link prediction-specific methods have been introduced \cite{zhang2023page,khemani2024review}.
Wang et al. \cite{wang2021pairwise} present PLNLP by jointly using the representations learned by a GNN, distance encoding, etc. Neo-GNN \cite{yun2021neo} weighted aggregates the link prediction scores obtained by heuristics and a GNN. 
NBFNet \cite{zhu2021neural} generalizes traditional path-based link heuristics into a path formulation. Singh et al. \cite{singh2021edge} show that adding a set of edges to the graph as a pre-processing step can improve the performance of link prediction models. PermGNN \cite{roy2021adversarial} optimizes the neighborhood aggregator directly by link samples. Zhao et al. \cite{zhao2022learning} study counterfactual questions about link existence by causal inference. RelpNet \cite{wu2022relpnet} aggregates edge features along the structural interactions between two target nodes. Guo et al. \cite{guo2022linkless} propose cross-model distillation techniques for link prediction. Shang et al. \cite{shang2022improving} propose a negative link sampling method PbTRM based on a policy-based training method. Li et al. \cite{li2024graph} study the integration of large language models (LLMs) and prompt learning techniques with graphs, enhancing graph transfer capabilities across diverse tasks and domains.

\subsection{Limitation Analysis} \label{sec:case study}
We provide two basic insights into the application of GNNs in link prediction. Firstly, we show that aggregation-based GNNs lack the ability to learn NCN-dependent structural information for link prediction (Remark~\ref{prop:GNN_cn}). Secondly, we demonstrate that node embeddings can boost the performance of GNN-based link prediction models on dense graphs (Remark~\ref{prop:gnn_ne_strong} and Finding \ref{finding:denser}). These can serve as effective avenues to identify and interpret the limitations of existing link prediction methods.
To illustrate this, we present two case studies. 

\begin{figure*}[!ht]
  \centering
  \includegraphics[width=0.97\linewidth]{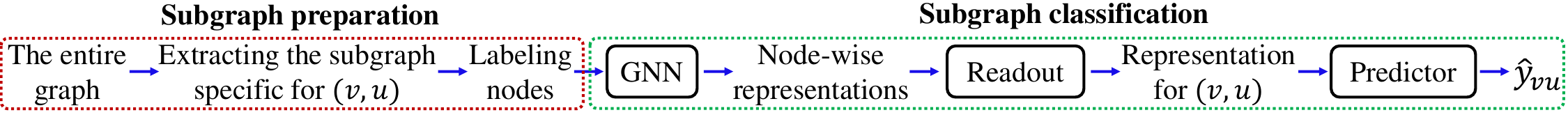}
  \caption{The algorithmic flow of SEAL-type link prediction methods.}
  \label{fig:seal}
\end{figure*}
\noindent\textbf{Case study 1. Can SEAL effectively learn NCN-dependent structural information?} 
SEAL-type methods have achieved the best performance on several link prediction datasets \cite{zhang2018link,li2020distance,teru2020inductive,yin2022algorithm}. SEAL \cite{zhang2018link} has proven that most link heuristics between two nodes can be computed approximately within an enclosing subgraph extracted specifically for that two nodes. As shown in Fig.~\ref{fig:seal}, most SEAL-type methods employ GNNs for graph representation learning, with the expectation that from such enclosing subgraphs, the GNN can learn the structural information equivalent to link heuristics including CN, AA, Katz, etc. However, whether this expectation holds true has not been thoroughly investigated in existing works. Herein we present a rough analysis to examine this issue.

First, the GNNs used in SEAL-type methods, e.g., DGCNN \cite{zhang2018end} in SEAL \cite{zhang2018link}, R-GCN \cite{schlichtkrull2018modeling} in GraiL \cite{teru2020inductive}, still belong to the type of aggregation-based GNNs. According to Remark~\ref{prop:GNN_cn}, these GNNs cannot effectively learn NCN-dependent structural information.

\begin{figure}
  \begin{center}
    \includegraphics[width=0.74\linewidth]{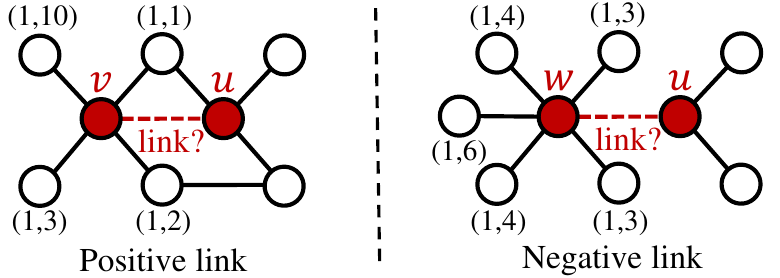}
  \end{center}
  \caption{Node labeling in SEAL-type methods. The left is a subgraph specific for a positive link sample and the right is a negative one. The labeling features are based on the SPDs from every node (here only show the first-order neighbors of node $v$ or $w$) to the target pair of nodes. For example, on the left, the node with the labeling $(1,10)$ indicates that the SPD from this node to node $v$ and $u$ is $1$ and $10$, respectively.}
  \label{fig:issues_label}
\end{figure}

Furthermore, we have noticed that SEAL-type methods typically use a labeling technique \cite{zhang2021labeling} to add labeling features to each node in the enclosing subgraph. 
The labeling features of each node describe the relationship of that node to the target two nodes. Fig. \ref{fig:issues_label} illustrates such a labeling method, where the labeling features of a node are the shortest path distances from the node to the target pair of nodes.  
The work \cite{zhang2021labeling} points out that the labeling features can help the GNN learn the structural information about the number of common neighbors. Their explanation is as follows. As shown in Fig. \ref{fig:issues_label}, for node $v$ and $u$, in the first iteration of the neighborhood aggregation of a GNN, only the common neighbors between node $v$ and $u$ will receive the labeling messages from both $v$ and $u$; then in the second iteration, the common neighbors will pass such messages back to both $v$ and $u$, which can encode the number of common neighbors into the representations of node $v$, $u$. 

However, the aforementioned explanation raises questions regarding its validity. In the second iteration, apart from the common neighbors, the non-common neighbors of node $v$ also pass their messages back to $v$. The messages from all neighbors of $v$ are then aggregated through a set-based pooling (e.g., MEAN or attention-based pooling as shown in Eq. \ref{eq:gcn_gat}). Such an aggregated result for node $v$ would wash out the distinguishable labels' information. We present an example to illustrate this. As shown in Fig.~\ref{fig:issues_label}, if the pooling method in a GNN is MEAN, then the aggregation of the labeling features of the neighbors of node $v$ would be equal to that of node $w$, i.e.,  $\mathrm{MEAN}(\{10,3,2,1\})=\mathrm{MEAN}(\{4,6,4,3,3\})$. This means that the distinct labeling features of the neighbors of a node are not effectively kept in the aggregated result. In other words, the aggregated results for node $v$ and $w$ in the positive and negative link samples become indistinguishable. Note that attention-based pooling in GNNs like GAT~\cite{gat2018graph} also suffers from the above limitation for the same reason as the analysis of Remark~\ref{prop:prop:gnn_num}. The same goes for node $u$. It should be noted that our example is merely for illustrative purposes. In practice, a GNN layer contains a series of complicated operations such as linear and non-linear transformations, dropout, residual connection, and others. The structural information in the labeling features could be partially kept in the learned representations.

\begin{table*}[t]
\caption{Results on OGB link prediction datasets. Higher scores indicate better performance, with the best results highlighted in bold. Herein, "HE" stands for Heuristic Encoding, "X" represents node attributes, and "NE" denotes Node Embedding.
}
\label{tab:results}
\begin{center}
    \normalsize
    \begin{tabular}{lcccc}
        \toprule
         & \multicolumn{1}{c}{\textbf{ogbl-ddi}} & \multicolumn{1}{c}{\textbf{ogbl-collab} } & \multicolumn{1}{c}{\textbf{ogbl-ppa}} &\multicolumn{1}{c}{\textbf{ogbl-citation2}}
        \\
        & \multicolumn{1}{c}{{Hits@20 (\%)}} & \multicolumn{1}{c}{{Hits@50 (\%)}} & \multicolumn{1}{c}{{Hits@100 (\%)}} &\multicolumn{1}{c}{{MRR (\%)}}
        \\
        \midrule
       
        MF~\cite{menon2011link}  &$13.68 \pm 4.75$& $38.86 \pm 0.49$ & $ 32.29\pm 0.94$ & $ 51.86\pm8.43 $
        \\
        FSSDNMF~\cite{chen2022link}  &$14.62 \pm 2.64$& $37.95 \pm 3.25$ & $ 34.15\pm 1.16$ & $ 54.71\pm8.73 $
        \\
        DeepWalk~\cite{perozzi2014deepwalk} &$ 26.42\pm 6.10$& $ 50.37\pm 0.34$ & $28.88 \pm 1.63$ & $60.11 \pm0.23 $
        \\
        Node2vec~\cite{grover2016node2vec} &$ 23.26\pm 2.09$& $48.88 \pm0.54 $ & $22.26 \pm 0.83$ & $ 61.41\pm0.11 $
        \\ 
        NodePiece ~\cite{galkin2021nodepiece} &$ 24.15\pm 3.04$& $ 47.88\pm 0.41$ & $22.85 \pm 0.94$ & $61.52 \pm2.91 $
        \\
        GraphSAGE~\cite{sage2017inductive} &$ 83.90\pm4.74$& $48.10 \pm 0.81$ & $16.55\pm2.40$ & $ 82.60\pm0.36 $
        \\
        GAT \cite{gat2018graph} &$95.38\pm0.94 $& $52.26\pm0.85 $ & $51.33 \pm2.16 $ & $ 83.17\pm0.54 $
        \\
        Neo-GNN \cite{yun2021neo} &$75.72\pm3.42 $& $55.31\pm0.53 $ & $49.13 \pm0.60 $ & $ 87.26\pm1.84 $
        \\
        PLNLP~\cite{wang2021pairwise} &$90.88 \pm 3.13 $& $ 52.92\pm 0.98 $ & $ 32.38 \pm 2.58 $ & $84.92 \pm 0.29  $
        \\ 
        NBFNet~\cite{zhu2021neural} &$18.14 \pm 2.12$& $51.15 \pm 1.38$ & $23.96 \pm 2.03$ & $74.91 \pm 2.37$
        \\
        SEAL~\cite{zhang2018link} &$ 30.56\pm3.86 $& $ 54.71\pm 0.79 $ & $ 48.80\pm4.56 $ & $87.67 \pm0.32 $ 
         \\
         DEGNN~\cite{li2020distance} & $26.63\pm6.42$ & $ 53.74\pm0.45 $ &$36.48\pm5.38$ & $ 60.30\pm0.81 $  
        \\
        SIEG \cite{ai2022structure} &$ 31.95\pm3.93 $& $ 55.35\pm0.52 $ & $ 53.35\pm1.39$ & $\mathbf{ 89.87 \pm 0.10} $  
        \\
        \midrule
        HE
         & $21.95 \pm 0.08$ & ${53.03 \pm 0.29} $ &$49.22\pm0.06$ & $ {81.91 \pm 0.05 }$  
        \\
         GCN(X) & $12.21 \pm 3.16$ & ${50.76 \pm 1.08} $ &$12.23\pm0.47$ & $ {80.60 \pm 0.04 }$  
        \\
         GCN(NE) &$\mathbf{95.49 \pm 0.73} $& ${50.17 \pm 0.56} $ & $ \mathbf{63.54\pm1.21} $ & $ 81.50 \pm 2.01 $  
        \\
        GCN(X+NE) &$94.42 \pm 0.63 $& ${53.08 \pm 0.46} $ & $ 60.68\pm3.52 $ & $ 83.85 \pm 0.03 $  
        \\
        GCN(X)+HE &$25.78 \pm 5.38 $& $\mathbf{56.11 \pm 0.64} $ & $ 49.68\pm0.39 $ & $88.63 \pm 0.05 $  
        \\
        GCN(NE)+HE &$91.11 \pm 3.60 $& ${54.21 \pm 0.07} $ & $ 61.02\pm2.51$ & $ 85.57 \pm 0.19 $  
        \\
        GCN(X+NE)+HE &$89.65 \pm 2.32 $& ${55.70 \pm 0.24} $ & $ 59.94\pm4.62$ & $ 86.31 \pm 0.12 $  
        \\
        \bottomrule\\[-8pt]
    \end{tabular}
\end{center}
\end{table*}
\begin{figure*}[ht!]
\begin{center}
\centerline{\includegraphics[width=0.9\textwidth]{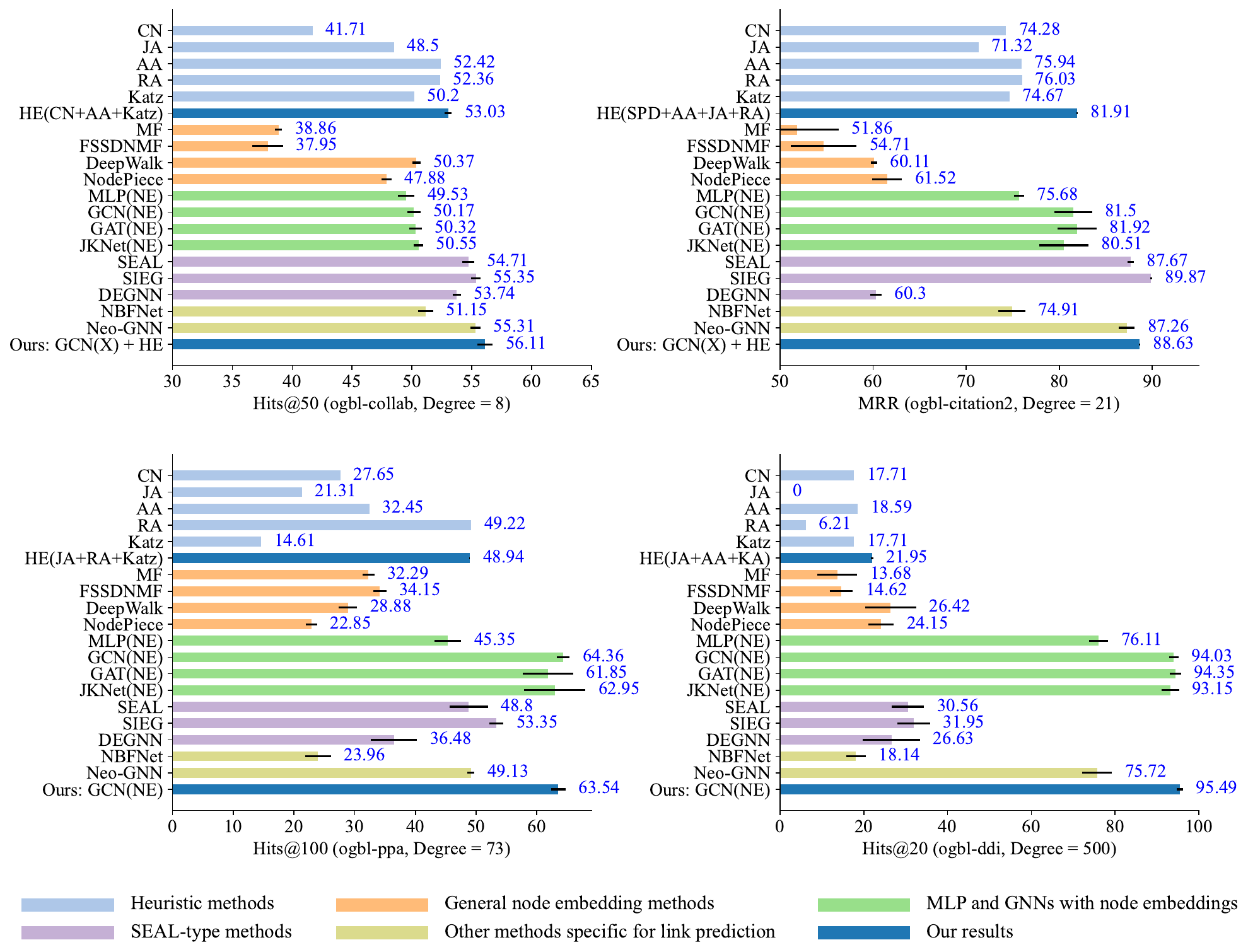}}
\caption{Results of different methods for link prediction on four OGB datasets. For MLP and general GNNs, we present their results obtained by utilizing node embeddings, considering the dominant performance of node embeddings as shown in Fig. \ref{fig:hexne}.}
\label{fig:baselines}
\end{center}
\end{figure*}

\begin{figure*}[!ht]
  \centering
  \includegraphics[width=0.7\linewidth]{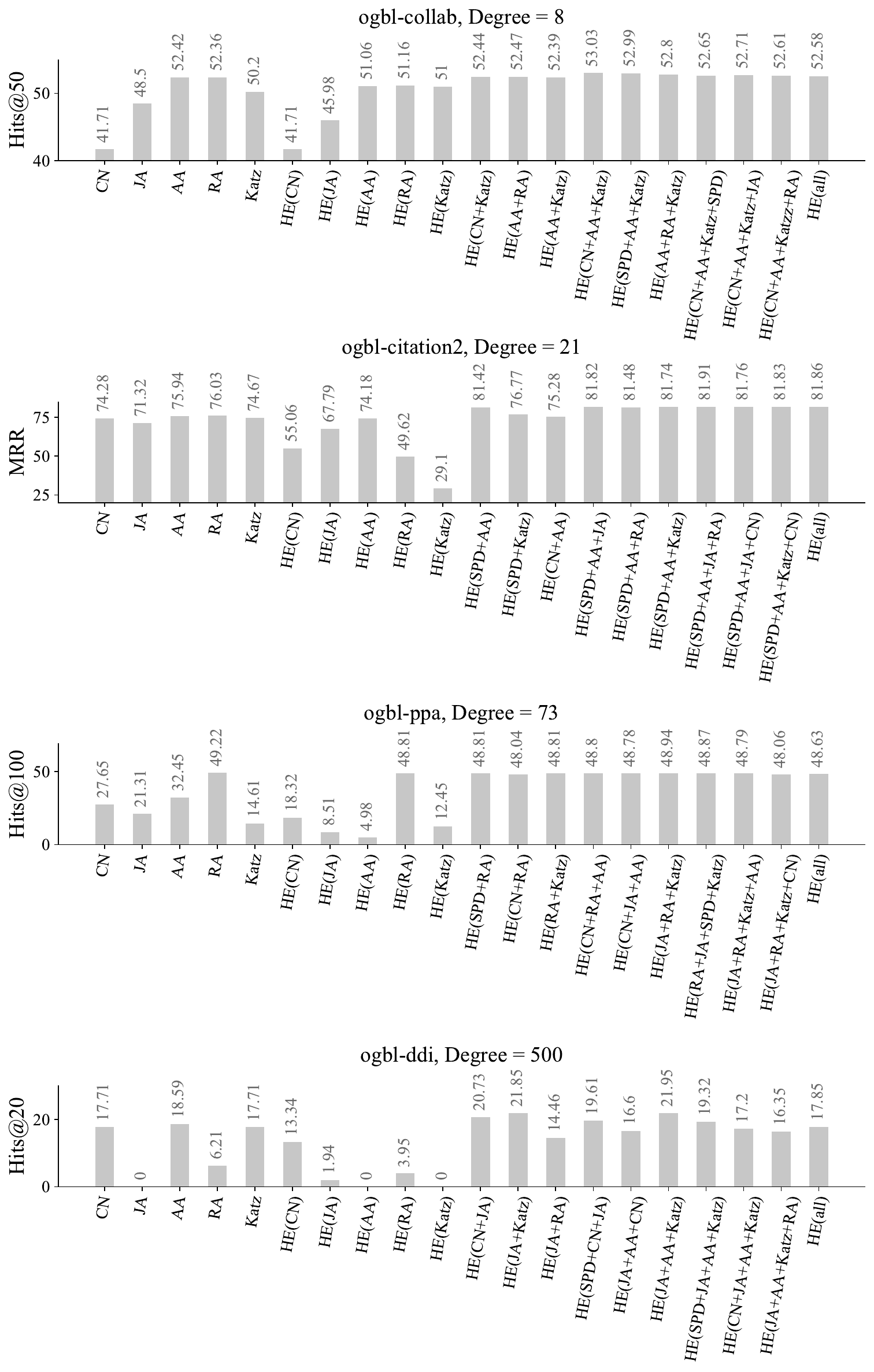}
  \caption{Results of HE (Heuristic Encoding) through Algorithm~\ref{alg:GNNLP}, where we encode various combinations of NCN-dependent heuristics. We can find that the best HE is often achieved by encoding a select number of heuristics rather than all heuristics.}
  \label{fig:HE}
\end{figure*}

Although SEAL-type methods cannot effectively learn structural information related to the number of common neighbors, we highlight that these methods are powerful for link prediction. Such methods transform the pair-specific link prediction problem into a graph-level classification task. Compared to models like Algorithm~\ref{alg:GNNLP} that only combine the representations of two target nodes, SEAL-type methods take advantage of the representations of not only two target nodes but also their neighboring nodes in the enclosing subgraph, enabling the model to consider more information of the surrounding environment of the candidate link.

\noindent\textbf{Case study 2. NBFNet lacks the algorithmic ability to leverage node embeddings.} 
NBFNet \cite{zhu2021neural} is a model specifically developed for link prediction. Differently from GNN-based link prediction methods like Algorithm~\ref{alg:GNNLP} and SEAL-type methods, NBFNet generalizes traditional link heuristics such as Katz index~\cite{katz1953new}, Personalized PageRank \cite{langville2011google} into a general formulation and approximates such formulation using a special network. Unlike aggregation-based GNNs that propagate and aggregate node-wise representations, NBFNet is designed to train edge-wise representations. The model architecture of NBFNet makes it hardly consider node-wise information. This would make NBFNet lack the algorithmic ability to train powerful node embeddings and may lead to non-competitive link prediction performance on dense graphs, considering the strong performance of the GNN only using node embeddings on dense graphs as shown in Fig.~\ref{fig:hexne}.

\subsection{Further analysis of experimental results}

We expand our limitation analysis of existing link prediction methods by examining their experimental performance on four OGB benchmark datasets. The results are presented in Table~\ref{tab:results} and Fig.~\ref{fig:baselines}. In the interest of brevity, our analysis focuses on several main types of link prediction methods.

First, as shown in Fig.~\ref{fig:baselines}, the performance of heuristic methods is not stable across the four datasets. For example, RA performs best on ogbl-ppa but second worst on ogbl-ddi. These results are consistent with the research of \cite{kovacs2019network} which indicates that many link heuristics are designed for specific applications and may perform well only on those specific graphs. Moreover, the unstable performance of every single heuristic confirms the need of combining multiple heuristics in link prediction, as shown in our Algorithm \ref{alg:GNNLP}. 

We also report the heuristic encoding results (Fig.~\ref{fig:HE}) obtained through Algorithm~\ref{alg:GNNLP} when only employing heuristic encoding. We find that it is not always true that the more the heuristics used, the better the performance of HE. In contrast, encoding a certain number of heuristics can yield the best results, whereas encoding too many heuristics would degrade performance. This may be due to optimization challenges, such as certain heuristics introducing strong correlations or biases that lead the model to converge to suboptimal local minima.

The node embedding methods based on relative distance encoding (DeepWalk \cite{perozzi2014deepwalk}, NodePiece~\cite{galkin2021nodepiece}) perform slightly better than those based on matrix factorization (MF \cite{menon2011link}, FSSDNMF \cite{chen2022link}). Nevertheless, all these methods fall short compared to other methods. This could be attributed to the limitations of such methods, e.g., reliance solely on the adjacency matrix or unsupervised learning without link samples. This also underscores the critical role of link samples in supervising the training of node embeddings for link prediction.

Fig.~\ref{fig:baselines} also presents the results of MLP and general GNNs (GCN \cite{kipf2017semi}, GAT \cite{gat2018graph} and JKNet \cite{xu2018representation}) that use node embeddings only. MLP(NE) performs much worse than GNNs, demonstrating the significance of the neighborhood aggregation of GNNs in training node embeddings, considering that MLP updates each node's representation independently of other nodes. 
Furthermore, GCN(NE) and GAT(NE) perform comparably, indicating that the expressiveness of GCN is sufficient for learning node embedding. The similar performance of GCN and GAT empirically supports our analyses in Remark \ref{prop:prop:gnn_num} and \ref{prop:GNN_cn}, where we point out that the attention mechanism (e.g., GAT) cannot address the inherent issue of GNNs in learning structural information related to the number of each node's neighbors and of common neighbors between two nodes.

In Fig.~\ref{fig:baselines}, SEAL-type methods show state-of-the-art performance. Especially, SIEG~\cite{ai2022structure} achieves the best results on two sparse graphs, i.e., ogbl-collab and ogbl-citation2 with graph degrees of $8$ and $21$, respectively. However, they perform worse than general GNNs with node embeddings (GCN(NE) \cite{kipf2017semi}, GAT(NE) \cite{gat2018graph} and JKNet(NE) \cite{xu2018representation}) on two dense graphs, i.e., ogbl-ppa and ogbl-ddi. This discrepancy in the performance of SEAL-type methods could be attributed to the algorithmic challenge of training node embeddings using subgraphs. Unlike general GNNs, the algorithm of SEAL-type methods limits each node to perceive other nodes within the subgraph rather than the entire graph, thereby restricting the information flow between nodes and potentially reducing the efficiency of learning node embeddings.

Besides, Fig.~\ref{fig:baselines} shows two link prediction-specific methods, namely NBFNet \cite{zhu2021neural} and Neo-GNN \cite{yun2021neo}. NBFNet underperforms on four datasets, which aligns with the limitations identified in Section \ref{sec:case study}. Neo-GNN predicts link likelihood by combining the scores obtained by heuristic methods and the result produced by a GNN. It performs on par with the state-of-the-art SEAL-type methods on two sparse graphs (ogbl-collab and ogbl-citation2).

Lastly, our GNN(X)+HE based on Algorithm \ref{alg:GNNLP} performs better than SEAL-type methods on ogbl-collab, supporting our limitation analysis of SEAL-type methods in Section \ref{sec:case study}, i.e., such methods could not effectively learn the information equivalent to NCN-dependent heuristics. 

It should be noted that our focus does not lie in developing solutions to these limitations of the existing methods, as this goes beyond the scope of our main goal. Nevertheless, these identified issues could pave the way for future research.

\section{Implication for Practical Applications}
\label{sec:Implication}

This study carries significant implications for real-world link prediction applications. A particular emphasis is the selection of appropriate solutions tailored to the graph degree.

For link prediction on sparse graphs, the performance of various methods in our experiments highlights the important role of NCN-dependent information. Approaches that can leverage such information, such as SEAL-type methods and Neo-GNN, generally outperform those that cannot. In practical scenarios, both SEAL-type methods and GNNs with heuristic encoding could yield satisfactory performance. Additionally, traditional machine learning models like MLP incorporating multiple heuristic encodings serve as viable alternatives, which could achieve comparable performance to the top-performing methods, while offering faster processing times. This is particularly advantageous for tasks such as recommender systems that demand rapid model response.

For link prediction on dense graphs, the contribution of node embeddings becomes dominant. Simple GNNs like GCN\cite{kipf2017semi} with the incorporation of trainable node embeddings can outperform most existing methods, rendering such a solution an optimal choice. However, this does not mean that the model using node embeddings will certainly perform better than that only using node features, especially in practical applications where careful feature engineering guided by domain knowledge is conducted. Besides, the use of trainable node embeddings remains limitations in the inductive setting \cite{teru2020inductive}, where new nodes are added to the graph, and the model together with all node embeddings may need to be retrained. In such cases, the methods that do not involve the training of node embeddings may offer more practical suitability.

\section{Limitations} 
\label{sec:Limitations}
This paper primarily explores several fundamental issues in link prediction methods, particularly in GNNs. It does not seek to introduce novel model architectures. Some analyses in this paper are provided in the form of examples and may lack rigorous mathematical proofs. 

\section{Conclusion}
\label{sec:Conclusion}

Link prediction stands as a pivotal task within the realm of graph applications. Our exploration into this domain reveals noteworthy variations in the performance of various link prediction methods across different graphs, with a significant dependence on graph degrees. Notably, on dense graphs, we observe that straightforward GNNs, like GCN, exhibit superior link prediction performance compared to many models that are developed specifically for link prediction. In contrast, on sparse graphs, the simple common-neighbor method often outshines GNN-based approaches. Understanding and interpreting these performance fluctuations, which is the objective of this work, is imperative, serving as a compass for refining existing methodologies and establishing a foundation for the development of more effective link prediction algorithms.

In addition, this work brings suggestions to practitioners in link prediction. Specifically, on sparse graphs, either SEAL-type methods or GNNs plus heuristic encoding can yield satisfactory performance. On dense graphs, GNN with node embeddings is an ideal choice in the transductive setting. For inductive learning, methods that do not involve the training of node embeddings may be more suitable.

\bibliography{main}
\bibliographystyle{IEEEtran}


\end{document}